\documentclass[aps,prd,twocolumn,showpacs,groupedaddress,nofootinbib]{revtex4}
\usepackage{graphicx}

\begin{document}

\centerline{~~~~~~~~~~~~~~~~~~~~~~~~~~~~~~~~~~~~~~~~~~~~~~~~~~~~~~~~~~~~~~~~~~~~~~~~
Published in PHYSICAL REVIEW D72, 034504 (2005).}

\title{Phase diagram of QCD at finite temperature and chemical potential from lattice simulations with dynamical Wilson quarks}

\author{He-Sheng Chen}
\address{Department of Physics, Zhongshan (Sun Yat-Sen) University,
Guangzhou 510275, China\\
and Department of Physics, Yangzhou University, Yangzhou 225009,
China}

\author{Xiang-Qian Luo}
\thanks{Corresponding author. Email address: stslxq@zsu.edu.cn}
\address{
CCAST (World Laboratory), P.O. Box 8730,
Beijing 100080, China\\
Department of Physics, Zhongshan (Sun Yat-Sen) University,
Guangzhou 510275, China}
\thanks{Mailing address.}

\date{\today}

\begin{abstract}
We present the first results for lattice QCD at finite temperature
$T$ and chemical potential $\mu$ with four flavors of Wilson
quarks. The calculations are performed using the imaginary
chemical potential method at $\kappa=0$, 0.001, 0.15, 0.165, 0.17
and 0.25, where $\kappa$ is the hopping parameter, related to the
bare quark mass $m$ and lattice spacing $a$ by $\kappa=1/(2ma+8)$.
Such a method allows us to do large scale Monte Carlo simulations
at imaginary chemical potential $\mu=i \mu_I$. By analytic
continuation of the data with $\mu_I < \pi T/3$ to real values of
the chemical potential, we expect at each $\kappa\in [0,
\kappa_{chiral}]$, a phase transition line on the $(\mu, T)$
plane, in a region relevant to the search for quark gluon plasma
in heavy-ion collision experiments. The transition is first order
at small or large quark mass, and becomes a crossover at
intermediate quark mass.
\end{abstract}

\pacs{12.38.Gc, 11.10.Wx, 11.15.Ha, 12.38.Mh}

\maketitle

\section{Introduction}

  QCD predicts a transition between
   hadronic matter (quark confinement) to quark-gluon plasma (quark deconfinement)
 at sufficient high temperature $T$ and small chemical $\mu$.
This matter might have existed in early universe immediately after the big bang.
The main purpose of heavy-ion collision experiments  at LHC (CERN), SPS, and RHIC (BNL) is to recreate such an
   environment. At large $\mu$ and lower $T$, several QCD-inspired models predict the existence of a
   color-superconductivity phase, which might be relevant to neutron star or quark star physics.
  Therefore, it is of great significance to study the QCD phase structure at larger $\mu$.
   Because QCD is still strongly coupled at criticality, perturbative methods
   do not apply. Lattice gauge theory (LGT) is the most reliable tool for investigating the phase transition
   from first principles.

In the continuum, the thermodynamics is described by the grand partition function
      \begin{eqnarray}
        \label{partition}
          Z(\mu,T)={\rm Tr} ~ e^{- \left( \hat{H}-\mu \hat{N_q} \right)/T} ,
       \end{eqnarray}
where $\hat{H}$ is the Hamiltonian,  $\mu$ is the chemical potential, and $\hat{N_q}=\int d^3x {\bar \psi} \psi$
is the quark number operator.

In the Lagrangian formulation of SU(3), LGT at finite $\mu$,
the effective fermionic action is complex and traditional Monte Carlo (MC) techniques
with importance sampling do not work.
The recent years have seen enormous efforts\cite{Muroya:2003qs,Katz:2003up,Lombardo:2004uy}
on solving the complex action problem,
and some very interesting information\cite{Fodor:2002hs} on the phase diagram for
QCD with Kogut-Susskind (KS) fermions at large $T$ and small $\mu$
has been obtained.
The KS approach to lattice fermions thins the fermionic degrees of freedom in naive fermions,
but it does not completely solve the species doubling problem.
It preserves the partial chiral symmetry,  but it breaks the flavor symmetry.
One staggered flavor corresponds to
four flavors in reality and
the fermionic determinant is replaced by the fourth root.
Such a replacement is mathematically un-proven\cite{Neuberger:2004be} and it might lead to the locality problem
in numerical simulations\cite{Bunk:2004br}.

Wilson's approach to lattice fermions\cite{Wilson:1974sk}
avoids the species doubling and preserves the flavor symmetry,
but it explicitly breaks the chiral symmetry,
one of the most important symmetries of the original theory.
Non-perturbative fine-tuning of the bare fermion mass has to be done,
in order to define the chiral limit.
There have been some MC simulations of LGT with Wilson fermions at finite temperature,
but no numerical investigation at finite chemical potential.
In Ref. \cite{Luo:2004mc}, the QCD phase structure on the $(\mu,T)$ plane
was investigated using Hamiltonian lattice QCD with Wilson fermions at strong coupling.
For $N_f/N_c<1$, with $N_c=3$ being the number of colors and $N_f$ being the number of flavors,
a tricritical point is found, separating the first order
and second order chiral phase transitions.

Neuberger's overlap fermion approach\cite{Neuberger:1997fp} solves the species doubling problem and preserves
the chiral symmetry and flavor symmetry, by introducing exponentially decaying non-local terms in the action.
However, the computational costs\cite{Fodor:2003bh} for dynamical overlap fermions
are typically two orders of magnitude heavier than for the Wilson or KS formulations.
This is  certainly beyond the current computer capacity.

In this paper, we perform the first MC simulations of Lagrangian
LGT with four flavors of dynamical Wilson quarks at finite temperature
and imaginary chemical potential $i\mu_I$, and study the phase
structure in the $(\mu, T,\kappa)$ space, by analytical
continuation to the real chemical potential.

\section{LATTICE FORMULATION}

The basic idea of lattice gauge theory\cite{Wilson:1974sk}, as
proposed by K. Wilson in 1974, is to replace continuous space-time
by a discrete grid with lattice spacing $a$. Gluons live on links
$U_j(x)=e^{-ig\int_x^{x+ \hat{j}a} dx' A_{j}(x')}$, and quarks
live on lattice sites. The continuum
Yang-Mills action $S_g=\int d^4x ~{\rm Tr} ~ F_{jl}(x)F_{jl}(x)/2$
is replaced by
\begin{eqnarray}
S_g=-{\beta \over 6} \sum_p {\rm Tr} (U_{p} +U_{p}^{\dagger}-2),
\label{gauge}
\end{eqnarray}
where $\beta=6/g^2$, and  $U_p$ is the ordered product of link variables $U$ around an elementary
plaquette.  The continuum quark action
$S_f=\int d^4x ~\bar{\psi}^{cont} (x) (\gamma_{j}D_{j}+m) \psi^{cont}(x)$  is replaced by
\begin{eqnarray}
S_f &=& \sum_{x,y} {\bar \psi}(x) M_{x,y}\psi(y).
\label{quark}
\end{eqnarray}
For Wilson fermions, the quark field $\psi$ on the lattice is
related to the continuum one $\psi^{cont}$ by $\psi=\psi^{cont}
\sqrt{a^3/(2\kappa)}$ with $\kappa=1/(2ma+8)$ the hopping
parameter. $M$ is the fermionic matrix:
\begin{eqnarray}
    M_{x,y}=\delta_{x,y} &-&
\kappa \sum_{j=1}^{4} \bigg[ (1-\gamma_{j})U_{j}(x)\delta_{x,y-\hat{j}}
\nonumber \\
&+&
(1+\gamma_{j})U_{j}^{\dagger}(x-\hat{j})\delta_{x,y+\hat{j}} \bigg] .
\end{eqnarray}
If we consider $N_f$ degenerate flavors of quarks, the partition function is
       \begin{eqnarray}
        \label{QCD_partition}
          Z &= &\int [dU][d\bar\psi][d\psi]e^{-S_g-S_f}
\nonumber \\
             &=& \int [dU] \left({\rm Det} M[U]\right)^{N_f} e^{-S_g}  .
       \end{eqnarray}

In the Lagrangian formulation of LGT, following Ref.
\cite{Hasenfratz:1983ba}, the chemical potential is introduced by
replacing the link variables in the temporal direction in
  fermion action with:
  \begin{eqnarray}
        \label{replacements}
          U_4(x)\rightarrow e^{a\mu} U_4(x),  ~~
          U^{\dag}_4(x)\rightarrow e^{-a\mu} U^{\dag}_4(x).
     \end{eqnarray}
The fermionic action is reduced to the continuum one when $a \to 0$.
However, the effective fermionic action in the partition function
becomes complex, and forbids MC simulation with importance sampling.
 Several revised methods, e.g., improved reweighting\cite{Fodor:2001au} and
  imaginary chemical potential\cite{Lombardo:1999cz,Azcoiti:2004ri} methods,
were proposed to simulate QCD with KS fermions at finite $\mu$.

Lattice QCD at imaginary chemical potential does not suffer the complex action problem.
In this paper, we will apply this method to the MC study
of  the phase structure of  Wilson fermions. We will measure the expectation of the Polyakov loop,
chiral condensate and their susceptibilities.

The Polyakov loop (Wilson line) is defined as
    \begin{eqnarray}
    \label{Polyakovloop}
    P(\vec{x})  =   {\rm Tr} \left[\prod_{t=0}^{N_t -1}{U_4(\vec{x},t)} \right] ,
\end{eqnarray}
where $N_t$ is the number of lattice sites in the temporal direction.
$P(\vec{x})$ is used to scale the interaction strength  between quarks.
In pure gauge theory, non-zero  $P(\vec{x})$ is a signal of quark deconfinement.
In practice, we measure
its expectation value over configurations generated in MC simulations with the probability distribution
$\left({\rm Det} M[U]\right)^{N_f} e^{-S_g}/Z$.

The chiral condensate is defined as
    \begin{eqnarray}
    \label{p_bar_p}
    \langle \bar\psi \psi \rangle &=& \frac{1}{Z}\int [dU] [d\bar\psi][d\psi]\bar\psi \psi e^{-S_g-S_f}
\nonumber \\
             &=&{1\over Z VN_t} \int [dU] {\rm Tr} \left( M^{-1}[U] \right)   \left({\rm Det} M[U]\right)^{N_f} e^{-S_g}  ,
\nonumber \\
   \end{eqnarray}
where $V$ is the spatial lattice volume. With suitable subtraction,
it serves as the order parameter of chiral-symmetry breaking
in the non-perturbatively defined chiral limit, $\kappa=\kappa_{chiral}$ where the pion is massless.

Strictly speaking, if the dynamical quarks play a role,
$P(\vec{x})$ is no longer an order parameter for deconfinement;
$\langle \bar\psi \psi \rangle$ is no longer an order parameter
for spontaneous chiral-symmetry breaking, if $\kappa \ne
\kappa_{chiral}$. However, when the system is at criticality, in
particular for the first order transition, one should observe
sharp changes in these quantities. Of course, this method is very
rough. At the transition point, there will be a peak in the
susceptibility, because the fluctuation of physical
  quantities is very strong. Therefore, the susceptibility will provide more useful information about the
transition. The susceptibility of a quantity
      \begin{eqnarray}
       O=\frac{1}{VN_t}\sum_{x,t}O(x)
      \end{eqnarray}
is defined as
    \begin{eqnarray}
      \label{replacements2}
       \chi = V N_t \langle (O^2 - \langle O \rangle^2 ) \rangle .
\end{eqnarray}
At criticality, the maximum value of $\chi$ behaves as $\chi_{max}
\propto V^{\alpha}$, with $\alpha$ the critical exponent. If
$\alpha=0$, the transition is just a crossover; If $0<\alpha<1$,
it is a second order phase transition; If $\alpha=1$, it is a
first order phase transition, accompanying the double peak
structure in the histogram of the quantity $O$ and flip-flops
between the two states in the MC history.

\section{PHASE DIAGRAM ON THE $(\mu_I, T)$ PLANE}

Many years ago, Roberge and Weiss (RW)\cite{Roberge:1986mm} made the first analytical study of
the phase structure of a gauge theory with fermions. Replacing
$\mu$ by  $i \mu_I$ ($\mu_I$ being a real  number) and introducing  $\theta = \mu_I/T$,
the grand partition function (\ref{partition}) has the behavior \cite{Roberge:1986mm}
   \begin{eqnarray}
        \label{periodic_Z}
        Z(\theta)=Z(\theta + 2\pi) .
     \end{eqnarray}
If the gauge group is SU$(N_c)$, the exact period is $2\pi/N_c$.
It implies the theory has a $Z_3$ symmetry for $N_c=3$, which is
clearly an artifact of imaginary $\mu$ and unphysical.

\begin{figure} [htbp]
\begin{center}
\includegraphics[totalheight=2.0in]{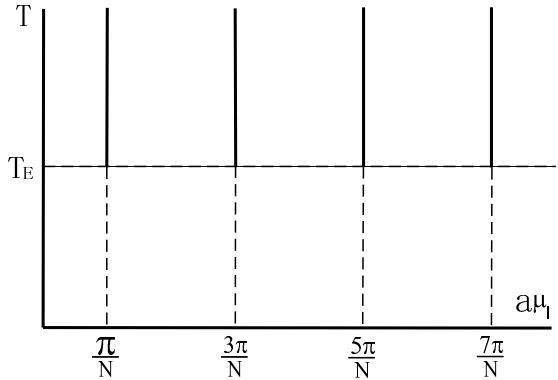}
\end{center}
\caption{Schematic phase diagram of QCD on the $(\mu_I, T)$ plane,
suggested by Roberge and Weiss. For $T \ge T_E$, there are first
order phase transitions at $a\mu_I=2\pi (k + 1/2)/N$,
characterized by discontinuity in the Polyakov loop. Here
$N=N_cN_t$ and $N_t a=1/T$. [There is no phase transition across
the dashed lines; they just illustrate the location of $T_E$ or
$a\mu_I=2\pi (k + 1/2)/N$].} \label{fig1}
\end{figure}

\begin{figure} [htbp]
\begin{center}
\includegraphics[totalheight=2.0in]{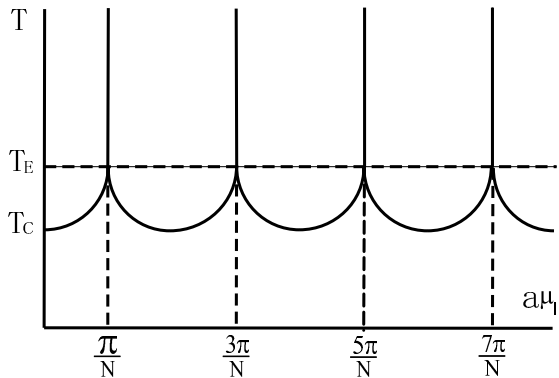}
\end{center}
\caption{Schematic phase diagram of QCD on the $(\mu_I, T)$ plane,
suggested by lattice MC study.} \label{fig2}
\end{figure}

Figure \ref{fig1} shows the phase diagram of QCD at imaginary chemical potential,
suggested by Roberge and Weiss\cite{Roberge:1986mm}.
Below some $T_E$, there is no phase transition. Above $T_E$ and at $\theta = 2\pi (k + 1/2)/N_c$,
with $k=0, 1, 2, ...$,
there are first order RW phase transitions between different $Z_3$ sectors,
characterized by the appearance of  discontinuity in the Polyakov loop.
However, later MC simulations with KS fermions\cite{deForcrand:2002ci,D'Elia:2002gd}
suggest a phase diagram as shown in Fig. \ref{fig2}, i.e.,
there are additional chiral/deconfinement transition lines for  $T_C \le T \le T_E$, with $T_C$ being the
critical temperature at $\mu=0$.
However, only the critical line lies within $\theta \in [0,\pi/N_c)$ and $T \in [T_C, T_E)$ is relevant for the
analytical continuation to the real chemical potential in the physical case.
I.e., the imaginary chemical potential method works only for $\mu_I/T<\pi/N_c$.

\section{MC SIMULATIONS}

In this section, we will present the first results for QCD with four
flavors of Wilson fermions at finite $T$ and $i\mu_I$.

The $R$ algorithm\cite{Gottlieb:1987mq} was used. We modified the
MILC collaboration's public LGT code\cite{Milc} to simulate the
case of imaginary chemical potential. The simulations were done at
lattice size $V\times N_t =8^3 \times 4$ and hopping parameter
$\kappa=$0, 0.001, 0.15, 0.165 0.17, and 0.25. At some $a\mu_I$,
$\beta$ and $\kappa$ values, finite size scaling analysis was
performed on different lattices. There are 20 molecular steps with
size $\delta \tau = 0.02$ for each configuration. For each $\beta$
and $a\mu_I$, we generated at least 20,000 configurations, after
4000 warmups. 20 iterations are carried out between measurements.
Around the area where the thermodynamical observables change
rapidly, we raised the statistics at least two times. All
simulations were done on our PC clusters. The cluster with 20
Pentium III-500 CPUs\cite{Luo:2000iz,Luo:2002kr} was built in
2000, and has been upgraded to  60 CPUs, with 40 new AMD
Opteron-242 CPUs.

\begin{figure} [htbp]
\begin{center}
\includegraphics[totalheight=2.0in]{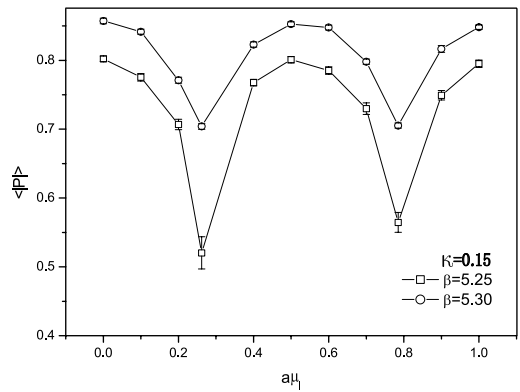}
\end{center}
\caption{Polyakov loop norm as a function of $a\mu_I$ at
$\kappa=0.15$ and two different $\beta$.} \label{fig3}
\end{figure}

\begin{figure} [htbp]
\begin{center}
\includegraphics[totalheight=2.0in]{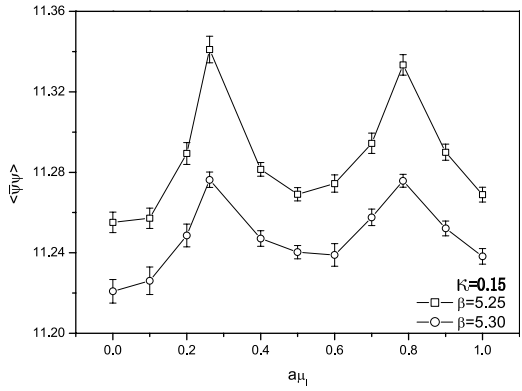}
\end{center}
\caption{Chiral condensate as a function of  $a\mu_I$ at
$\kappa=0.15$ and two different $\beta$.} \label{fig4}
\end{figure}

\begin{figure} [htbp]
\begin{center}
\includegraphics[totalheight=2.0in]{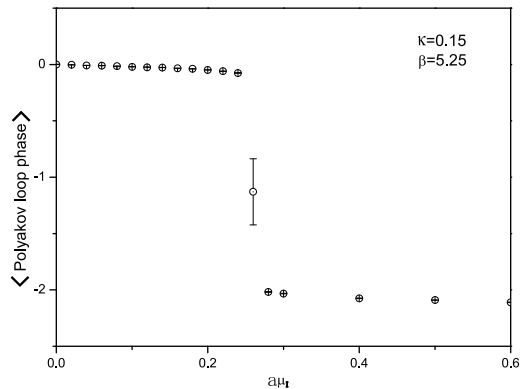}
\end{center}
\caption{Phase of the Polyakov loop as a function of $a\mu_I$ at
$\beta=5.25$ and $\kappa=0.15$.} \label{fig5}
\end{figure}

\begin{figure} [htbp]
\begin{center}
\includegraphics[totalheight=2.0in]{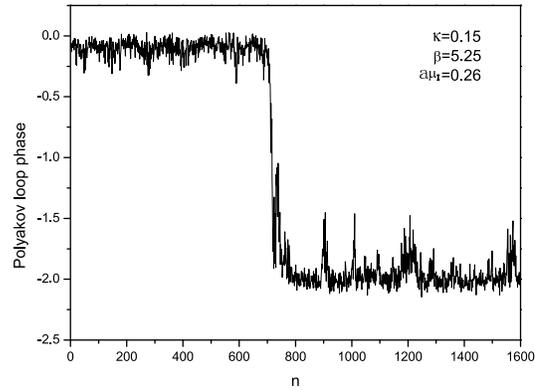}
\end{center}
\caption{MC history of the phase of the Polyakov loop at
 $a\mu_I=0.26$, $\beta=5.25$, and $\kappa=0.15$.} \label{fig6}
\end{figure}

We have more complete data at $\kappa=0.15$ than other $\kappa$
values. Figure \ref{fig3} plots the results for the Polyakov loop
norm $\langle \vert  P({\vec x}) \vert \rangle $ as a function of
$a\mu_I$ at two different and larger values of $\beta$
(corresponding to higher $T$).
 The data for  $\langle \bar\psi \psi\rangle$ are shown in Fig. \ref{fig4}. As one sees,
  these quantities are approximately periodic with
  period $\pi/6$, confirming the RW period for $a\mu_I$ to be $2\pi aT/3=2\pi/(3N_t)$.
Furthermore, at $a\mu_I =2(k +1/2)\pi/(3N_t)$, $k=0, 1, ...$,
there is a rapid change in the thermodynamical quantities,
indicating the RW phase transition at higher $T$ (larger $\beta$).
Figure \ref{fig5} is a more detailed scan for the phase of the
Polyakov loop at $\beta=5.25$. At $a\mu_I= 0.26$, it changes
rapidly. Figure \ref{fig6} plots the MC history of the phase of
the Polyakov loop. There is a clear signal for first order phase
transition at $a\mu_I=0.26$, confirming the existence of RW
transition between different $Z_3$ sectors at
$a\mu_I=\pi/(3N_t)=\pi/12$.

To locate the chiral/deconfinement phase transition line, we made
more detailed measurements of
  $\langle\vert P({\vec x}) \vert \rangle$, $\chi_{\vert P \vert}$, $\langle \bar\psi \psi\rangle$,
and $\chi_{\bar\psi \psi}$ for $a\mu_I < \pi/(3N_t)$.

\begin{figure} [htbp]
\begin{center}
\includegraphics[totalheight=2.0in]{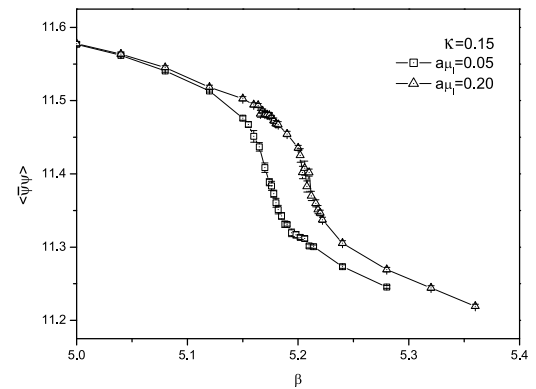}
\end{center}
\caption{Chiral condensate as a function of $\beta$ at
$\kappa=0.15$ and two different values of $a\mu_I$.} \label{fig7}
\end{figure}

\begin{figure} [htbp]
\begin{center}
\includegraphics[totalheight=2.0in]{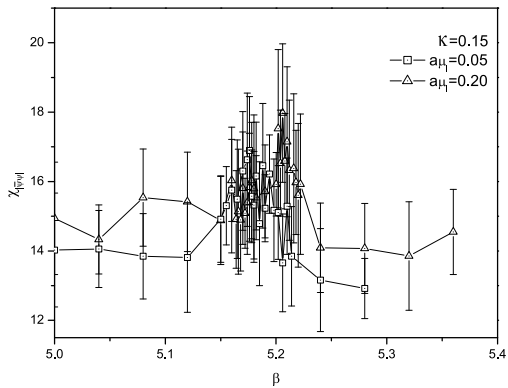}
\end{center}
\caption{Chiral susceptibility as a function of $\beta$ at
$\kappa=0.15$ and two different values of $a\mu_I$. } \label{fig8}
\end{figure}

\begin{figure} [htbp]
\begin{center}
\includegraphics[totalheight=2.0in]{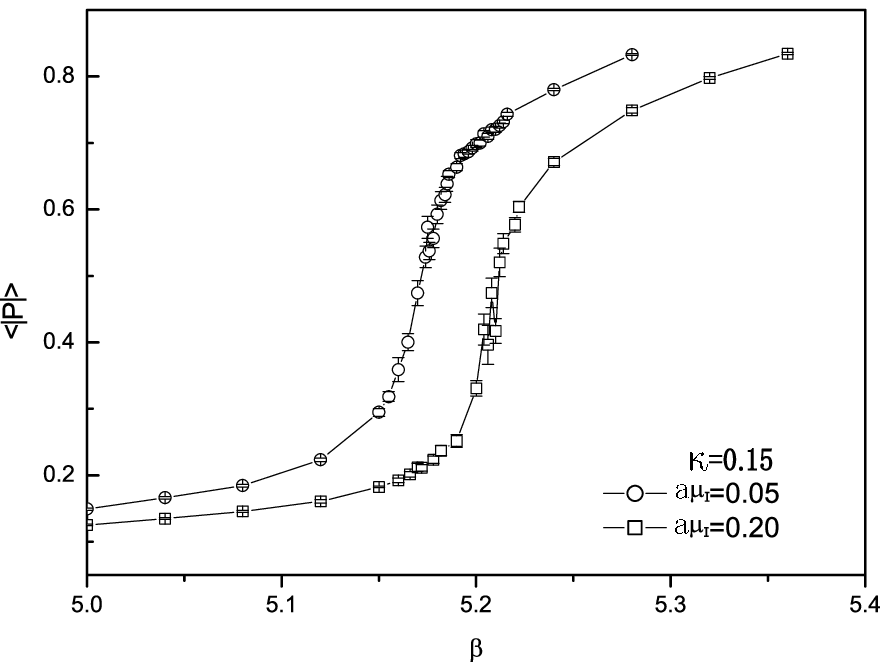}
\end{center}
\caption{Polyakov loop norm as a function of $\beta$ at
$\kappa=0.15$ and two different $a\mu_I$.} \label{fig9}
\end{figure}

\begin{figure} [htbp]
\begin{center}
\includegraphics[totalheight=2.0in]{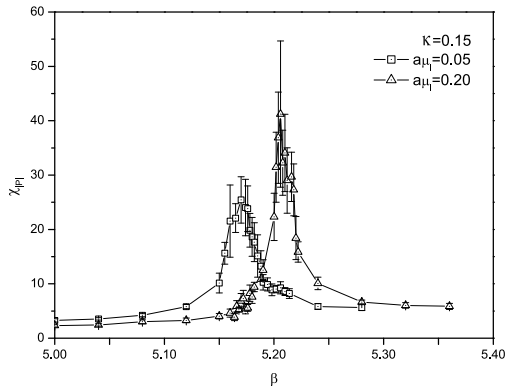}
\end{center}
\caption{Susceptibility of the Polyakov loop norm as a function of
$\beta$ at $\kappa=0.15$ and two different $a\mu_I$.}
\label{fig10}
\end{figure}

Figures \ref{fig7} and \ref{fig8} show respectively
$\langle\bar\psi\psi\rangle$ and $\chi_{\bar\psi \psi}$ versus
$\beta$ at $a\mu_I=0.05$ and 0.20 for $\kappa=0.15$;
 Figures \ref{fig9} and \ref{fig10}  show the results for $\langle \vert P \vert \rangle$ and
$\chi_{\vert P \vert}$.  In the rapid changing area, the Polyakov
loop and its susceptibility behave more singularly than the chiral
condensate and chiral susceptibility. From the position of the
peak in the susceptibilities, we determine the transition point. A
collection of transition points $(a\mu_I, \beta_C)$ is listed in
Tab. \ref{table1}. In Refs.
\cite{deForcrand:2002ci,D'Elia:2002gd}, it has generally been
argued that for small $a\mu_I$, the transition line $\beta_C
(a\mu_I)$ can be expressed as a Taylor series with even power of
$a\mu_I$. Due to limited data, the series is truncated to the
quadratic term. We use the least squares method to fit the data in
Tab. \ref{table1} for $\kappa=0.15$, and obtain an equation for
the transition line:
             \begin{eqnarray}
             \label{fits}
              \beta_C = 5.169(9) + 0.954(33) \left( a\mu_I \right)^2 + O\left(a^4\mu_I^4\right)
              ,
             \end{eqnarray}
with error bars coming from the fit.

\begin{table}
\begin{center}
\begin{tabular}{ccc}
  \hline
  $a\mu_I$ & & $\beta_C$ \\
  \hline
  0.00 & & 5.168(2)\\
  0.05 & & 5.170(2)\\
  0.10 & & 5.180(2)\\
  0.14 & & 5.187(2)\\
  0.18 & & 5.200(2)\\
  0.20 & & 5.206(2)\\
  0.22 & & 5.217(2)\\
\hline
\end{tabular}
\end{center}
\caption{Collection of transition points for $\kappa=0.15$,
determined by locating the peak of the susceptibilities, with
error bars coming from the scan precision.} \label{table1}
\end{table}

\begin{figure} [htbp]
\begin{center}
\includegraphics[totalheight=2.0in]{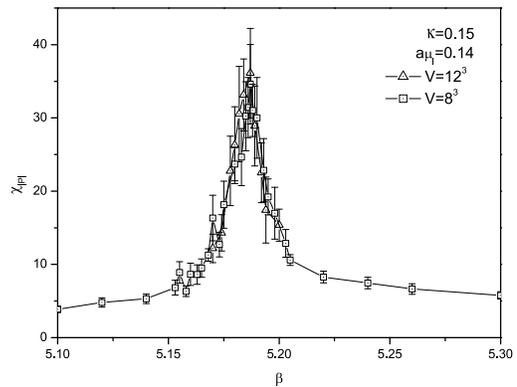}
\end{center}
\caption{Susceptibility of the Polyakov loop norm as a function of
$\beta$ at $a\mu_I = 0.14$ and $\kappa=0.15$ for different spatial
volumes $V=8^3$ and $12^3$.}
 \label{Sus8-12}
\end{figure}

\begin{figure} [htbp]
\begin{center}
\includegraphics[totalheight=2.0in]{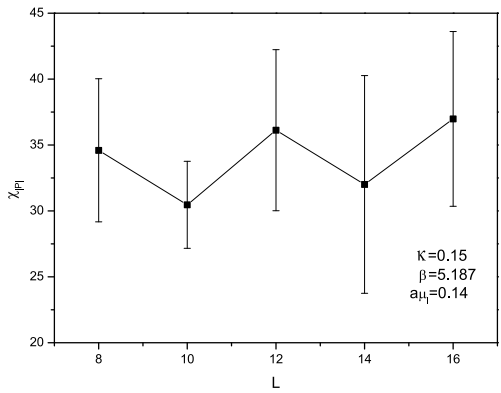}
\end{center}
\caption{Susceptibility of the Polyakov loop norm as a function of
the spacial extent $L$ at $a\mu_I= 0.14$, $\beta=5.187$ and
$\kappa=0.15$.}
 \label{Sus-vol}
\end{figure}

Nevertheless, at this $\kappa$, there is no obvious double peak
structure in the histograms of thermodynamical observables. To
determine the nature of the transition, one has to do a finite
size study. Let us take the transition point at $a\mu_I=0.14$ and
$\beta=5.187$ in Tab. \ref{table1} as an example. Figure
\ref{Sus8-12} compares $\chi_{\vert P \vert}$ for spacial volumes
$V=L^3=8^3$ and $12^3$ around the transition point (with $\Delta
\beta=0.002$). Within error bars, the locations and heights of the
peaks are consistent. At this $\beta$, we did the longest
simulations on $8^3\times 4$, $10^3\times 4$, $12^3\times 4$,
$14^3\times 4$, and $16^3\times 4$ lattices, with statistics at
least four times higher than other non-transition points. As shown
in Fig. \ref{Sus-vol}, $\chi_{\vert P \vert}$ does not increase as
$L$. This implies that the transition point at $a\mu_I=0.14$ and
$\beta=5.187$ for $\kappa=0.15$ is just a crossover.

\begin{figure} [htbp]
\begin{center}
\includegraphics[totalheight=2.0in]{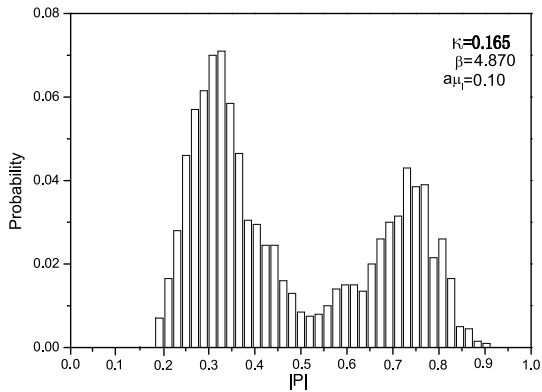}
\end{center}
\caption{Histogram of the Polyakov loop norm at and $a\mu_I =
0.1$, $\beta = 4.87$ and $\kappa=0.165$.}
 \label{histogram}
\end{figure}

\begin{figure} [htbp]
\begin{center}
\includegraphics[totalheight=2.0in]{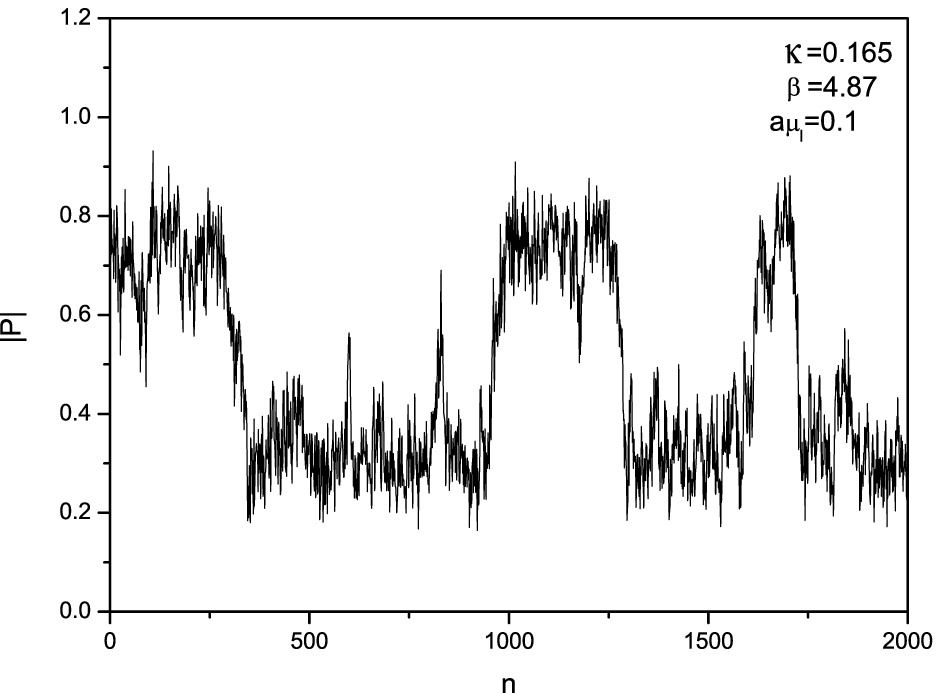}
\end{center}
\caption{MC history of the Polyakov loop norm at $a\mu_I = 0.1$,
$\beta =4.87$, and $\kappa=0.165$.}
 \label{simulation}
\end{figure}

\begin{figure} [htbp]
\begin{center}
\includegraphics[totalheight=2.0in]{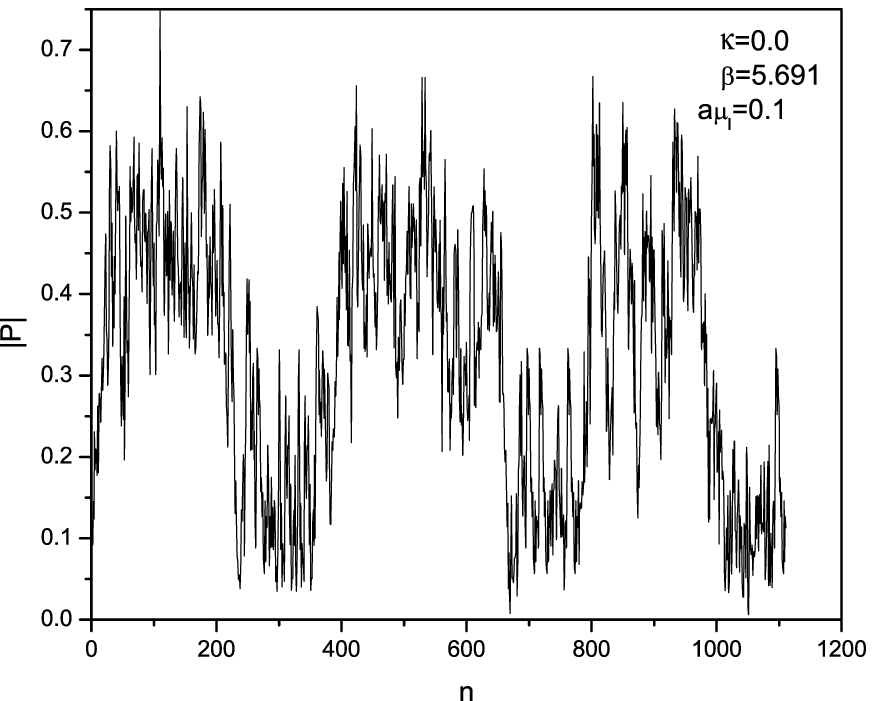}
\end{center}
\caption{MC history of the Polyakov loop norm at $a\mu_I = 0.1$,
$\beta = 5.691$, and $\kappa=0$.}
 \label{simulation0}
\end{figure}

\begin{figure} [htbp]
\begin{center}
\includegraphics[totalheight=2.0in]{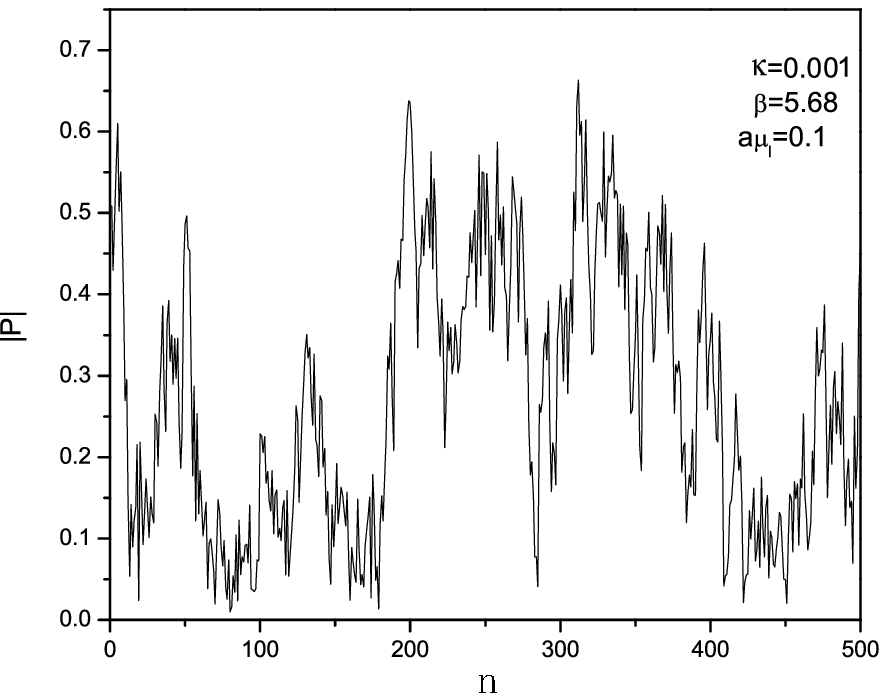}
\end{center}
\caption{MC history of the Polyakov loop norm at $a\mu_I = 0.1$,
$\beta = 5.68$, and $\kappa=0.001$.}
 \label{simulation0.001}
\end{figure}

\begin{figure} [htbp]
\begin{center}
\includegraphics[totalheight=2.0in]{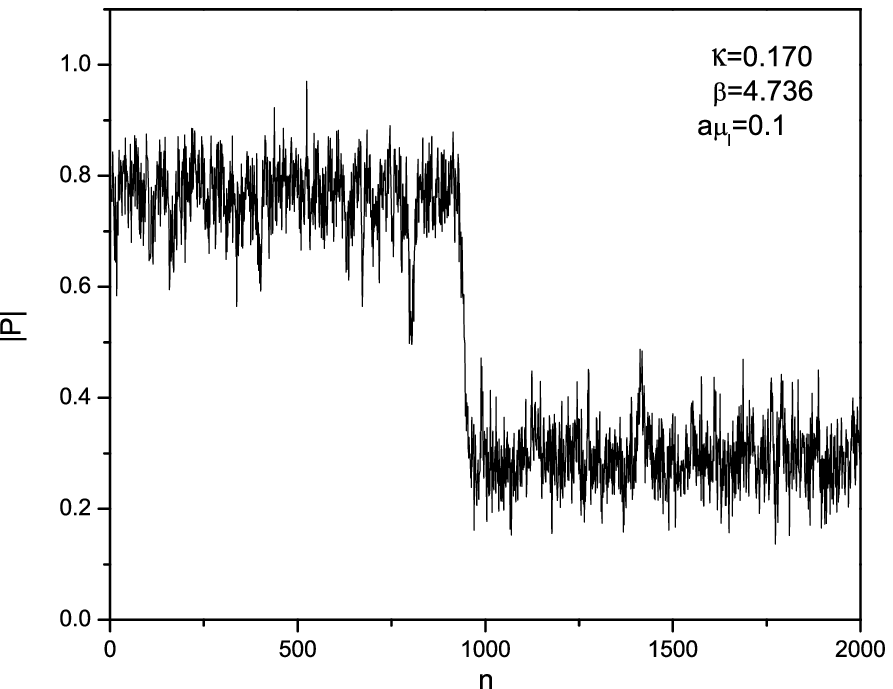}
\end{center}
\caption{MC history of the Polyakov loop norm at $a\mu_I = 0.1$,
$\beta = 4.736$, and $\kappa=0.17$.}
 \label{simulation0.17}
\end{figure}

\begin{figure} [htbp]
\begin{center}
\includegraphics[totalheight=2.0in]{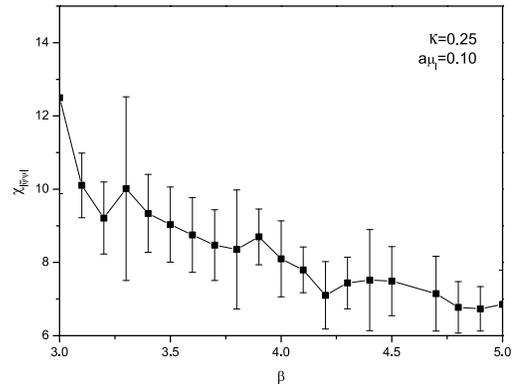}
\end{center}
\caption{Chiral condensate susceptibility as a function of $\beta$
at $a\mu_I=0.1$ and $\kappa=0.25$.} \label{psi-k0.25}
\end{figure}

\begin{figure} [htbp]
\begin{center}
\includegraphics[totalheight=2.0in]{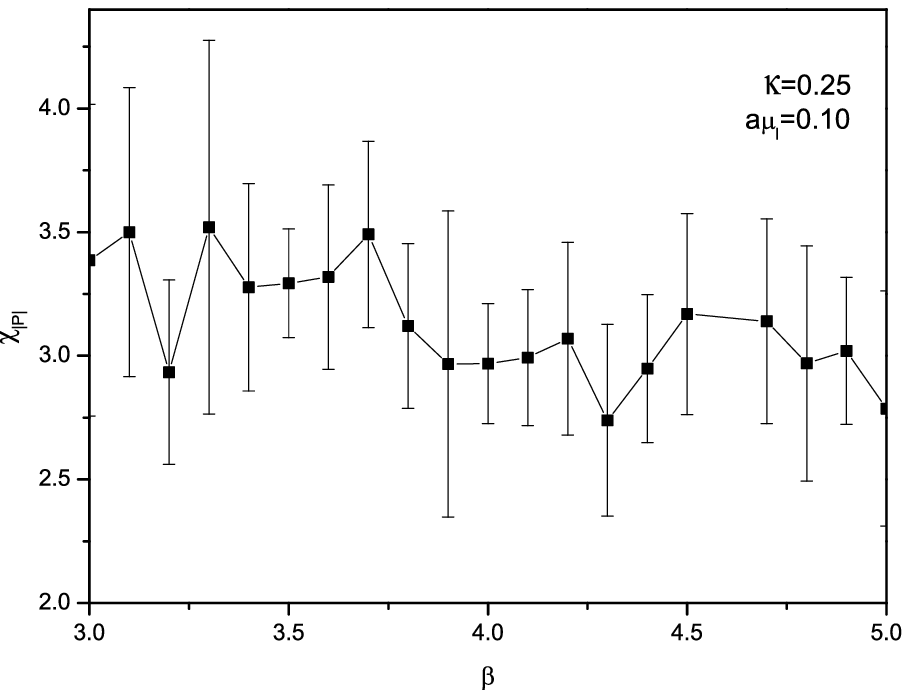}
\end{center}
\caption{Susceptibility of the Polyakov loop norm as a function of
$\beta$ at $a\mu_I=0.1$ and $\kappa=0.25$.} \label{pol-0.25}
\end{figure}

However, the situations at small or large $\kappa$ are very
different. Figure \ref{histogram} shows the histogram of the
Polyakov loop norm for  $\beta = 4.870$ and $a\mu_I=0.1$, at
$\kappa = 0.165$ which is closer to the chiral limit
$\kappa_{chiral}$. We observe a double peak structure. Figure
\ref{simulation} plots the history of the MC simulation at the
same parameters, with $\vert P \vert$ measured after warmups; We
observe the jumps of $\vert P \vert$ from one plateau to another.
The results for $\kappa=0$, 0.001 and 0.17 are similar, as shown
in Figs. \ref{simulation0}, \ref{simulation0.001} and
\ref{simulation0.17}. These indicate that the phase transitions at
$\kappa \in [0,\kappa_1]$ and $\kappa \in [\kappa_2,
\kappa_{chiral}]$ are of first order; Here $\kappa_1 \in (0.001,
0.15)$, $\kappa_2 \in (0.15, 0.165)$ and $\kappa_{chiral} \in
(0.17, 0.25)$. At $\kappa=0.25$, which should be the case when
$\kappa> \kappa_{chiral}$, Figs. \ref{psi-k0.25} and
\ref{pol-0.25} tell us that there is no phase transition.

\section{PHASE DIAGRAM ON THE $(\mu, T)$ PLANE: THE PHYSICAL CASE}

Replacing $\mu_I$ by $-i\mu$, we directly continue the transition
line (\ref{fits})  at $\kappa=0.15$ from imaginary chemical
potential to real chemical potential:

             \begin{eqnarray}
             \label{Extropolate}
              \beta_C = 5.169(9) - 0.954(33) \left(a\mu\right)^2 + O\left(a^4\mu^4\right) .
             \end{eqnarray}

\begin{figure} [htbp]
\begin{center}
\includegraphics[totalheight=2.0in]{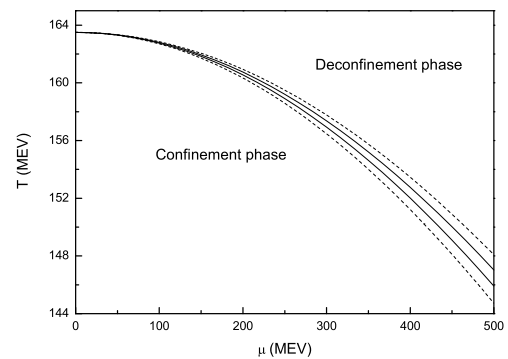}
\end{center}
\caption{Phase diagram on the $(\mu, T)$ plane for $\kappa=0.15$.
The area between the two solid lines is our result for Wilson quarks
with error band, derived from Eqs. (\ref{Extropolate}) and
(\ref{renormalization}). The area between the two dotted lines is
the result for KS fermions by D'Elia and Lombardo.
} \label{phaseTMU}
\end{figure}

In order to translate the lattice results into the physical units,
we use the renormalization group relation between the lattice
spacing $a$
 and $\beta$. The two loop perturbative  expression gives\cite{Cucchieri:2000hv}
             \begin{eqnarray}
              \label{renormalization}
              a\Lambda_L &= & \exp \bigg(-\frac{4\pi^2}{33-2N_f}\beta \\
\nonumber \\
                  & +& \frac{459-57N_f}{(33-2N_f)^2}\ln \left(\frac{8\pi^2}{33-2N_f}\beta\right) \bigg),
             \end{eqnarray}
where $\Lambda_L$ is the lattice QCD scale. In Ref.
\cite{Karsch:2000kv}, the phase transition was studied with
different kinds of fermion actions at $\mu = 0$ and consistent
results were obtained. Therefore we fix the lattice QCD scale by
the critical temperature $T_C=164$ MeV at $\mu = 0$
  with 4 flavors for KS fermions \cite{Cucchieri:2000hv}. The temperature $T$ is related
to $a$ and $N_t$ by $T=1/(aN_t)$. The critical line on the
$(\mu,T)$ plane is shown in Fig. \ref{phaseTMU}. For comparison,
the critical line for KS fermions\cite{D'Elia:2002gd} is also
shown. Within error bars, the results are consistent.

\section{DISCUSSIONS}

  In the preceding sections, we have studied the properties
 of the phase structure of four-flavor QCD on the $(\mu, T)$ plane,
 using the information obtained from MC simulations of LGT with Wilson fermions at imaginary chemical potential
$i\mu_I<i\pi T/3$.

The advantages of Wilson formulation have been mentioned in the
introduction: it is free of species doubling and there is one to
one correspondence between the flavors on the lattice and in the
continuum.

Our study suggests that QCD with four flavors of  Wilson quarks
experiences a first order phase  transition from the confinement
phase to the deconfinement phase at small or large quark mass.
However, at intermediate quark mass, the transition becomes a
crossover. The properties of the transition are similar to those of KS
fermions. This region is of interest for the present heavy-ion
collision experiments.

\begin{figure} [htbp]
\begin{center}
\includegraphics[totalheight=2.0in]{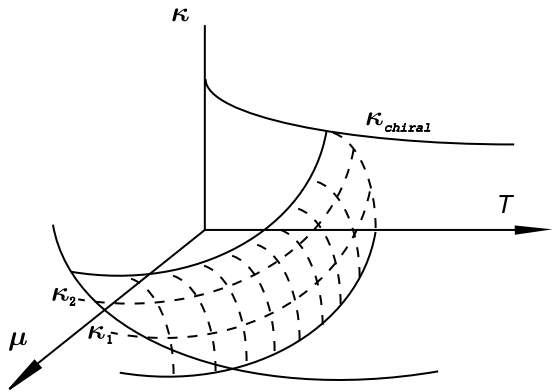}
\end{center}
 \caption{Expected phase diagram of lattice QCD with
four flavors of Wilson quarks in the $(\mu, T, \kappa)$ parameter
space. For $\kappa \in [0,\kappa_1]$ and $\kappa \in
[\kappa_2,\kappa_{chiral}]$, the phase transition is of first
order, and while for $\kappa \in (\kappa_1,\kappa_2)$, the
transition is a crossover.} \label{fig21}
\end{figure}

Figure \ref{fig21} is the expected phase diagram of lattice QCD
with Wilson fermions in  the $(\mu, T, \kappa)$ parameter space.
There is a surface $\kappa=\kappa_{chiral}$ where the pion becomes
massless. Above this surface, there is no phase transition, as
confirmed by our numerical simulations for $\kappa=0.25$.
Interesting physics is below this surface: at each $\kappa$, one
should see a phase structure similar to Fig. \ref{phaseTMU}. Of
course, the order of transition depends on the value of $\kappa$.
Due to heavy computational costs of simulations with dynamical
fermions, we did not do a comprehensive search for the exact
location of $\kappa_1$, $\kappa_2$ and $\kappa_{chiral}$. We
believe that $\kappa_1 \in (0.001, 0.15)$, $\kappa_2 \in (0.15,
0.165)$ and $\kappa_{chiral} \in (0.17, 0.25)$.

For lower temperature and larger chemical potential, all available
MC simulation techniques fail. In the Hamiltonian lattice
formulation, there has been successful analysis of the critical
behavior for strong coupling QCD
\cite{Gregory:1999pm,Luo:2000xi,Fang:2002rk} at $(\mu, T=0)$ and
on the $(\mu, T)$ plane\cite{Luo:2004mc}. To study the continuum
physics, new methods have to be developed. We hope to study these
issues, as well as the dependence of the phase structure on the
quark flavors, in the near future.

\acknowledgments

We thank M. D'Elia, C. DeTar, P. de Forcrand, E. Gregory, and M.
Lombardo for useful discussions. This work is supported by the Key
Project of National Science Foundation (10235040), Project of the
Chinese Academy of Sciences (KJCX2-SW-N10) and Key Project of
National Ministry of Eduction (105135) and Guangdong Ministry of
Education.


\begin{thebibliography}{99}

\bibitem{Muroya:2003qs}
S.~Muroya, A.~Nakamura, C.~Nonaka and T.~Takaishi,
Prog.\ Theor.\ Phys.\  {\bf 110}, 615 (2003), and refs. therein.

\bibitem{Katz:2003up}
S.~D.~Katz,
Nucl.\ Phys.\  B (Proc.\ Suppl.) \  {\bf 129}, 60 (2004), and refs. therein.


\bibitem{Lombardo:2004uy}
M.~P.~Lombardo,
Prog.\ Theor.\ Phys.\ Suppl.\  {\bf 153}, 26 (2004), and refs.
therein.

\bibitem{Fodor:2002hs}
Z.~Fodor and S.~D.~Katz,
JHEP {\bf 0203}, 014 (2002).

\bibitem{Neuberger:2004be}
H.~Neuberger,
Phys.\ Rev.\ D {\bf 70}, 097504 (2004).

\bibitem{Bunk:2004br}
B.~Bunk, M.~Della Morte, K.~Jansen and F.~Knechtli,
Nucl.\ Phys.\ B {\bf 697}, 343 (2004).




\bibitem{Wilson:1974sk}
K.~G.~Wilson,
Phys.\ Rev.\ D {\bf 10} (1974) 2445.

\bibitem{Luo:2004mc}
X.~Q.~Luo,
Phys. Rev. D {\bf 70}, 091504 (2004) (Rapid Commun.).


\bibitem{Neuberger:1997fp}
H.~Neuberger,
Phys.\ Lett.\ B {\bf 417}, 141 (1998)

\bibitem{Fodor:2003bh}
Z.~Fodor, S.~D.~Katz and K.~K.~Szabo,
JHEP {\bf 0408}, 003 (2004)




\bibitem{Hasenfratz:1983ba}
P.~Hasenfratz and F.~Karsch,
Phys.\ Lett.\ B {\bf 125}, 308 (1983).




\bibitem{Fodor:2001au}
Z.~Fodor and S.~D.~Katz,
Phys.\ Lett.\ B {\bf 534}, 87 (2002).

\bibitem{Lombardo:1999cz}
M.~P.~Lombardo,
Nucl. Phys. B (Proc. Suppl.)  {\bf 83}, 375 (2000).

\bibitem{Azcoiti:2004ri}
V.~Azcoiti, G.~Di Carlo, A.~Galante and V.~Laliena,
JHEP {\bf 0412}, 010 (2004).


\bibitem{Roberge:1986mm}
A.~Roberge and N.~Weiss,
Nucl.\ Phys.\ B {\bf 275}, 734 (1986).


\bibitem{deForcrand:2002ci}
P.~de Forcrand and O.~Philipsen,
Nucl.\ Phys.\ B {\bf 642}, 290 (2002).


\bibitem{D'Elia:2002gd}
M.~D'Elia and M.~P.~Lombardo,
Phys.\ Rev.\ D {\bf 67}, 014505 (2003).



\bibitem{Gottlieb:1987mq}
S.~A.~Gottlieb, W.~Liu, D.~Toussaint, R.~L.~Renken and R.~L.~Sugar,
Phys.\ Rev.\ D {\bf 35}, 2531 (1987).


\bibitem{Luo:2000iz}
X.~Q.~Luo, E.~B.~Gregory, J.~C.~Yang, Y.~L.~Wang, D.~Chang and Y.~Lin,
arXiv:hep-lat/0011090.

\bibitem{Luo:2002kr}
X.~Q.~Luo, E.~B.~Gregory, H.~J.~Xi, J.~C.~Yang, Y.~L.~Wang, Y.~Lin and H.~P.~Ying,
Nucl.\ Phys.\ B(Proc. Suppl.)  {\bf 106}, 1046 (2002).


\bibitem{Cucchieri:2000hv}
A.~Cucchieri and D.~Zwanziger,
Phys.\ Rev.\ D {\bf 65}, 014002 (2002).


\bibitem{Karsch:2000kv}
F.~Karsch, E.~Laermann and A.~Peikert,
Nucl.\ Phys.\ B {\bf 605}, 579 (2001).

\bibitem{Gregory:1999pm}
E.~B.~Gregory, S.~H.~Guo, H.~Kr\"oger and X.~Q.~Luo,
Phys.\ Rev.\ D {\bf 62},  054508 (2000).


\bibitem{Luo:2000xi}
X.~Q.~Luo, E.~B.~Gregory, S.~H.~Guo and H.~Kr\"oger,
hep-ph/0011120.


\bibitem{Fang:2002rk}
Y.~Fang and X.~Q.~Luo,
Phys. Rev. D {\bf 69}, 114501 (2004).


\bibitem{Milc} http://physics.utah.edu/$\sim$detar/milc/

\end{thebibliography}
\end{document}